\documentclass[ ,final] {aipproc}
\layoutstyle{6x9}

\begin{document}

\title{The Expanding Photosphere Method: Progress and Problems}


\author{J\'ozsef Vink\'o}{
  address={Department of Optics \& Quantum Electronics, University of Szeged, Hungary}
}

\author{Katalin Tak\'ats}{
  address={Department of Optics \& Quantum Electronics, University of Szeged, Hungary}
}


\begin{abstract}
Distances to well-observed Type II-P SNe are determined from an updated version of
the Expanding Photosphere Method (EPM), based on recent theoretical models. 
The new EPM distances show good agreement with other independent distances
to the host galaxies without any significant systematic bias, contrary to earlier
results in the literature. The accuracy of the method is comparable with that of the
distance measurements for Type Ia SNe. 
\end{abstract}

\maketitle

\section{Introduction}

Distance is one of the most fundamental quantities in astrophysics, and
it is especially true for supernovae. Type Ia SNe are thought to be
the most reliable distance indicators, even up to $z \sim 1.5$ redshift,
and they play major role in determining the expansion of the Universe
as well as the cosmic equation of state. On the other hand, accurate 
distances to SNe are crucial in understanding not only their physical 
properties, but also revealing their progenitor objects and the 
possible explosion mechanisms. 

The Expanding Photosphere Method (EPM) is a tool for measuring distances to
SNe that have large amount of ejected material \cite{kk74}. 
The concept of EPM is based on a few assumptions about the general physics of
the expanding ejecta. These are the followings:
\begin{enumerate}
\item{The expansion of the ejected material is spherically symmetric.}
\item{The ejecta is expanding homologously, i.e. $R(t) = v(R) \cdot (t - t_e)$, where
$R(t)$ is the time-dependent radius of a particular layer in the ejecta, $v(R)$ is the
(constant) expansion velocity of this layer and $t - t_e$ is the time elapsed since the moment
of explosion ($t_e$). }
\item{The ejecta is optically thick, i.e. there exists a layer where the optical depth
$\tau_\lambda \sim 1$. This layer is the ,,photosphere'' ($R_{phot}$). Because of the
expansion, the location of the photosphere moves inward the ejecta, so its velocity
 ($v_{phot}$) is decreasing with time. }
\item{The photosphere radiates as a blackbody, so the shape of the emergent
flux spectrum is Planckian with a well-defined effective temperature $T_{eff}$. 
However, the absolute flux value differs from that of the blackbody due to the
dominance of scattering opacity over true absorption in the ejecta. The deviation from the
blackbody can be described with a simple scaling, i.e. $F_\lambda = \zeta^2 \pi B_\lambda(T)$.
where $F_\lambda$ is the surface flux, $B_\lambda(T)$ is the Planck function and
$\zeta$ is the correction (or ``dilution'') factor.}
\end{enumerate}

These assumptions are most likely to be valid in Type II-P SNe. 
These eject a massive, hydrogen-rich envelope that remains optically thick for
$\sim 100$ days after explosion, and the emergent spectrum is indeed close 
to be Planckian. Thus, EPM is expected to work best for such SNe. 
  
Based on the assumptions, the instantaneous radius of the photosphere can be expressed as
$R_{phot} = v_{phot}(t) \cdot (t - t_e)$ (the radius of the progenitor is usually neglected). 
Meantime, the observed flux is $f_\lambda = \theta^2 \cdot \zeta^2 \pi B_\lambda (T)$, where 
$\theta = R_{phot} / D$
is the angular radius of the photosphere from distance $D$.  Combining these two equations,
one gets the basic equation of EPM \cite{hamuy01, leo1}:
\begin{equation}
t ~=~ t_e + D \cdot \left ( {\theta \over v_{phot}} \right ).
\end{equation}
Since $\theta$ and $v_{phot}$ can be determined from observations, $t_e$ and $D$ are
the only unknowns in Eq.1. These can be derived via least-squares fitting to the observed
quantities.

If the SNe under study are at high redshifts, the equations should be slightly modified \cite{schmidt94}.
The definition of the angular radius is connected with the angular distance $D_A$, while in the
expression of the observed flux the luminosity distance $D_L$ enters. At high $z$ 
$D_L = (1+z)^2 D_A$, so the angular radius of the photosphere can be inferred from 
\begin{equation}
\theta ~=~ {1 \over \zeta} \sqrt{{f_\lambda (1+z)} \over {\pi B_{\lambda'}(T)}},
\end{equation}
where $\lambda' = \lambda / (1+z)$. 

One particular advantage of EPM is that it does not
require initial calibration, i.e. a sample of objects with a priori known distances. However, 
the computation of the $\zeta$ correction factors needs detailed model atmospheres,
which makes the method essentially model-dependent. Currently, there are two independent sets
of model atmospheres of Type II-P SNe in the literature, which were used to compute correction factors as a
function of $T_{eff}$ \cite{eastman, dess1}. The former one was used in detailed studies
of SN~1999em (the most extensively studied SN II-P so far) that resulted in 
$D_{EPM} \approx 8 \pm 1$ Mpc \cite{hamuy01, leo1, elm1} . 
This is in significant disagreement with the Cepheid distance to the host galaxy NGC~1637 
being $D_{Cep} = 11.7 \pm 1$ Mpc \cite{leo3}. This problem has been solved in \cite{dess2} by using
a new set of correction factors based on the NLTE radiative transfer code CMFGEN
which gave $D_{EPM} = 11.5 \pm 1.0$ Mpc for SN~1999em.

\section{New EPM distances to SNe II-P}

\begin{figure}
\includegraphics[width=8cm]{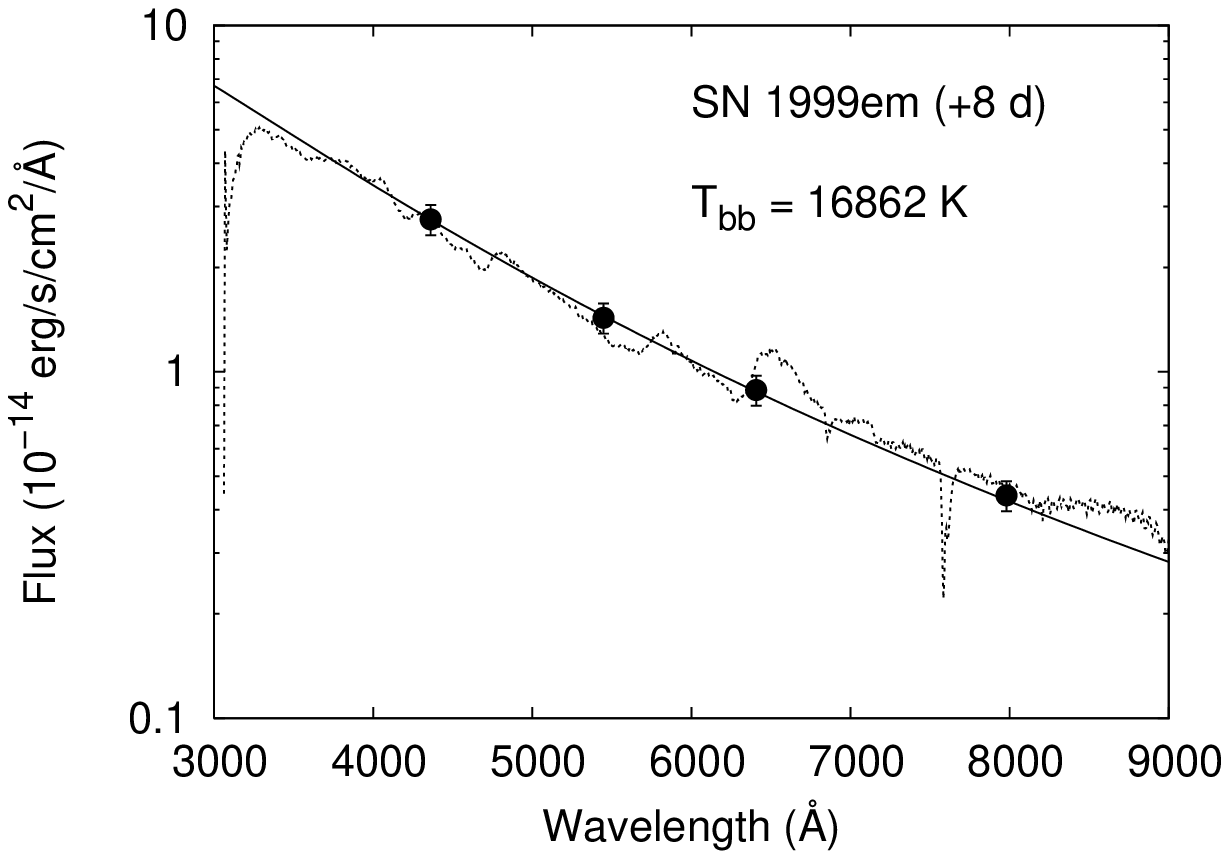}
\includegraphics[width=8cm]{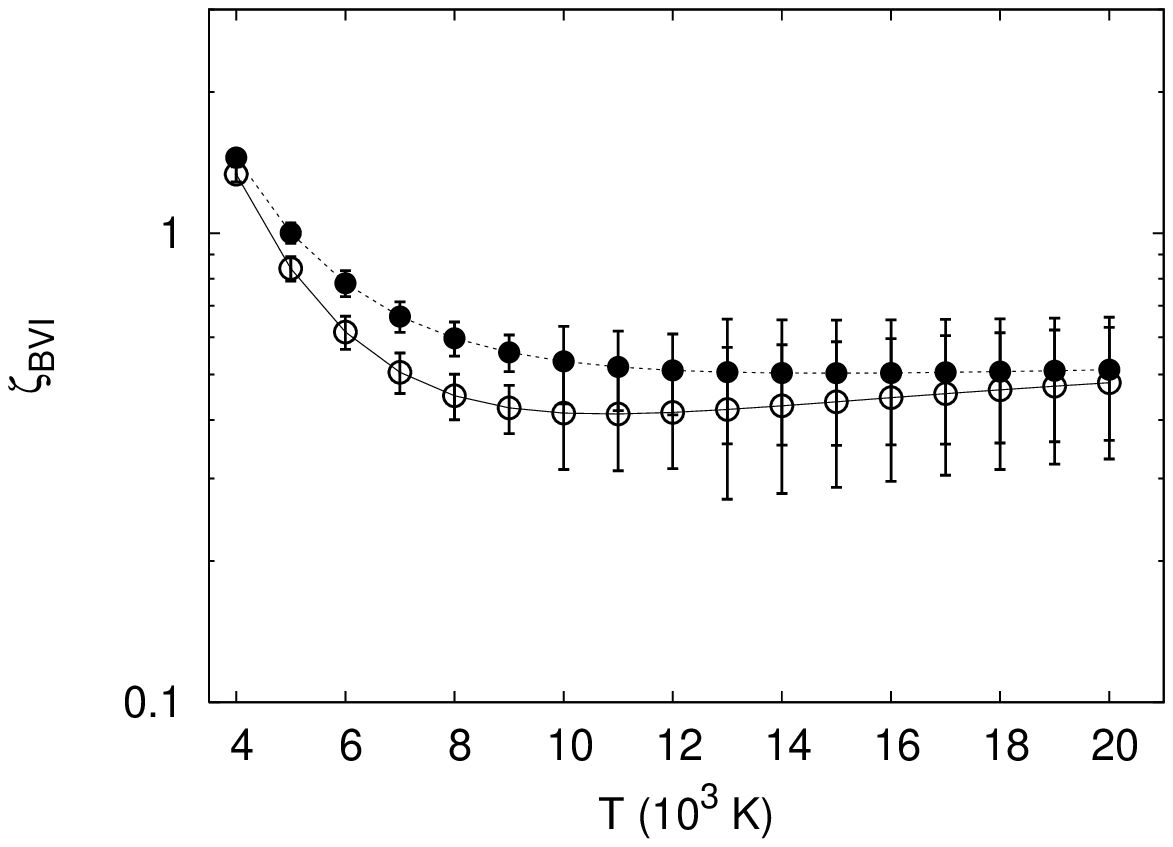}
\caption{Left panel: Result of fitting a blackbody (solid line) to broadband $BVI$ fluxes (filled symbols) of 
SN 1999em \cite{leo1}. The $R-$band flux is also in good agreement with the fitted blackbody. The flux-calibrated 
spectrum obtained simultaneously (dotted line)  is shown for comparison.  Right panel: The correction factor as
a function of $T_{eff}$ from \cite{dess1} (filled circles) and \cite{eastman} (open circles).}
\end{figure}

The method outlined above has been implemented in a new code that needs observed $BVRI$
light curves, radial velocities (determined from the absorption minima of certain spectral
features, see below) and reddening information (typically $E(B-V)$) as input.  As in any method
based on photometry, the magnitudes must be dereddened, but fortunately the results of EPM are 
quite insensitive to reddening errors, compared with other methods \cite{eastman}. 

At each epoch, the angular radius has been computed by a simultaneous fitting to the dereddened 
$B$, $V$ and $I$ fluxes, as described in \cite{hamuy01}. The corresponding effective temperature
has been derived by fitting a blackbody curve to the broadband fluxes converted from the dereddened
magnitudes. Our experience shows that the best results can be achieved by considering all optical+NIR
(i.e. $BVRI$) fluxes simultaneously. Earlier studies were sometimes limited to the usage of $B$ and
$V$ bands only, which may result in increased systematic errors due to the large deviation of the $B$-band
fluxes from the blackbody curve at later phases. The left panel of Fig.1 illustrates the optimum fitting of 
a blackbody to either photometric, or precisely calibrated spectroscopic fluxes. 

From the resulting $T_{eff}$, the correction factor $\zeta$ has been computed from the 
$\zeta_{BVI} (T)$ function of Dessart \& Hillier \cite{dess1} for data obtained less than 40 days
after explosion. For data measured between 40 - 60 days after explosion, the function given by Eastman 
et al. \cite{eastman} was applied. As noted above, the usage of the function of Dessart \& Hillier 
produces better distances, but their models are valid only during the first month after explosion, 
before the hydrogen starts to recombine.  The $\zeta_{BVI} (T)$ functions are plotted in the right panel of
Fig.1.

\begin{figure}
\includegraphics[width=8cm]{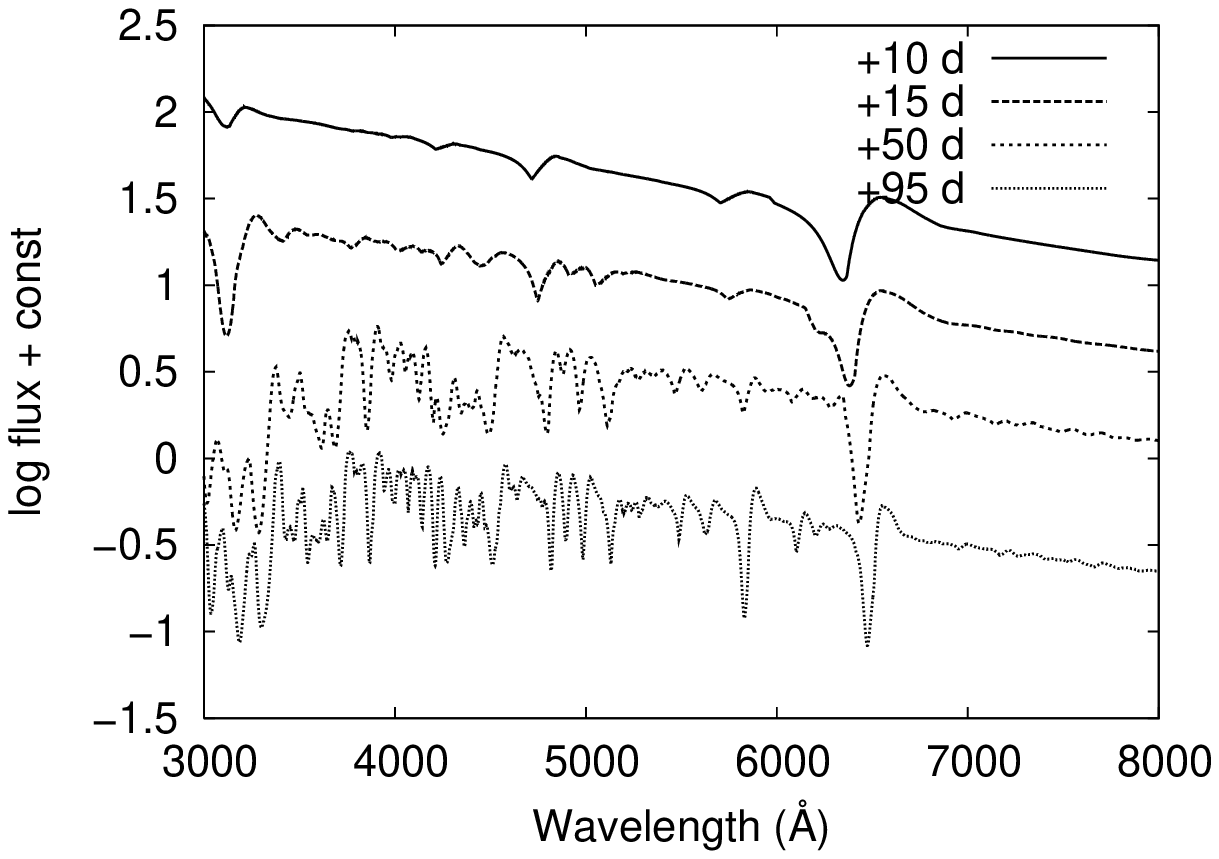}
\includegraphics[width=8cm]{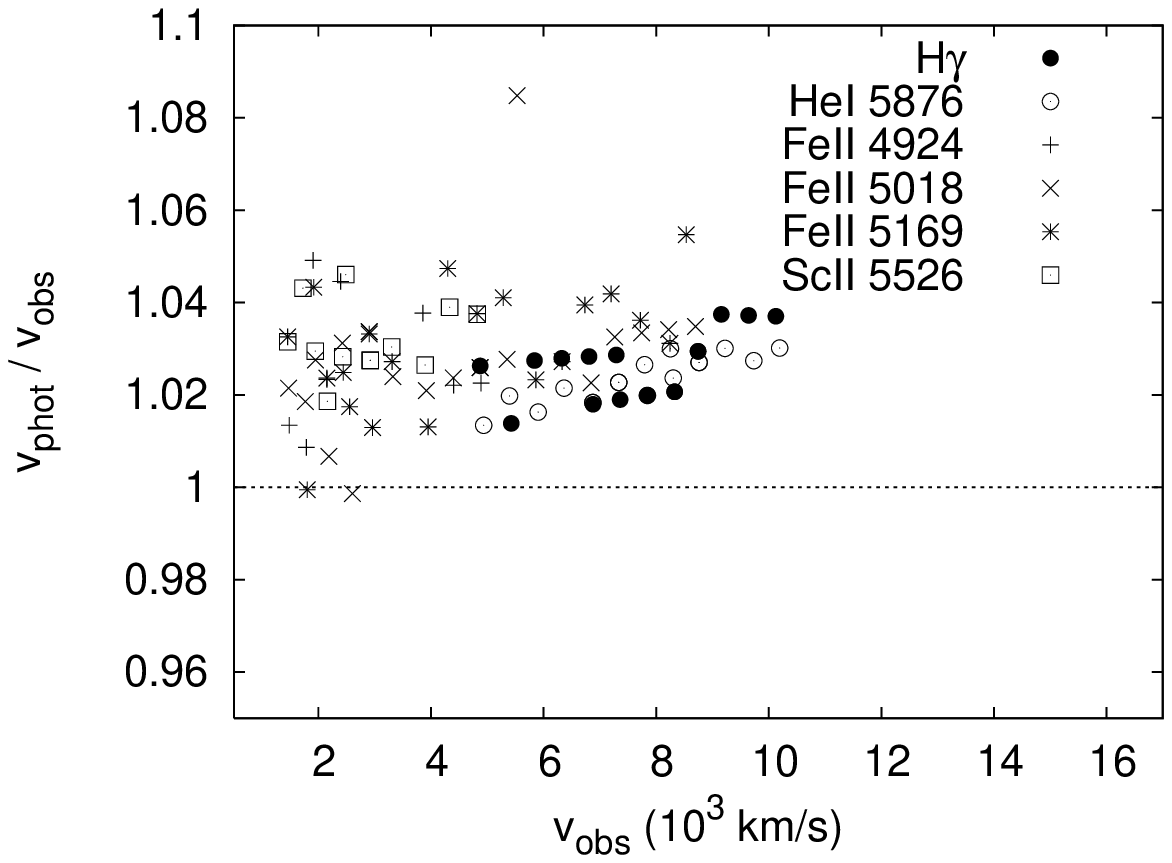}
\caption{Left panel: SYNOW model spectra of Type II-P SNe. The phase of each spectrum 
(expressed in days after explosion) is indicated. Right panel: The ratio of the true photospheric
velocity (an input parameter of a SYNOW model) to the ,,observed'' velocity (derived from the
absorption minimum of P Cygni lines) as a function of the ,,observed'' velocity. Different symbols
mean different photospheric lines indicated on the righ-hand side.}
\end{figure}

Beside the correction factors, the other important quantity is the photospheric velocity $v_{phot}$, 
because the resulting distance is very sensitive to the velocities that appear in the denominator in 
Eq.1. Thus, the problem of finding an optimum method to infer $v_{phot}$ from Type II-P SNe spectra 
has been addressed in several studies (see \cite{hamuy01, leo1, dess1}). 

We have studied this problem by computing model spectra with the parametrized spectral
synthesis code SYNOW \cite{baron1}. SYNOW computes the emergent spectrum in a
homologously expanding atmosphere assuming LTE and pure scattering line formation.
The input parameters are the velocity and the blackbody temperature at the 
photosphere ($v_{phot}$ and $T_{eff}$), the exponent of the atmospheric structure,  
the list of ions contributing to the spectral features, and the optical depth of one strong line for each ion.

Four sets of spectra have been defined corresponding
to phases $+10$, $+15$, $+50$ and $+95$ days after explosion, respectively. The list of ions
contained H, He I, Na I, Fe II, Sc II, Ti II and Ba II, because these ions are thought to be
responsible for the strongest lines in the optical \cite{leo1}. The input parameters except $v_{phot}$ 
were tuned to match real Type II-P SNe spectra. 
Then, several model spectra were synthesized with different input $v_{phot}$ for each phase. 
The left panel of Fig.2 shows representative spectra of all phases.  

The synthesized spectra were used to compute ``observed'' radial velocities by measuring 
the Doppler-shift of  the absorption minima of selected lines.  For P Cygni line profiles, this 
should give exactly  $v_{phot}$ if the line is isolated and optically thin. However, in reality 
the features in a SN spectrum are all blends and may not be optically thin. 
Therefore, $v_{obs}$ will differ from $v_{phot}$. 

In the right panel of Fig.2 the ratio of $v_{phot} / v_{obs}$ is plotted
as a function of $v_{obs}$ for the features shown. It is seen that in almost
all cases $v_{phot}$ is slightly underestimated. The explanation of such a phenomenon is 
discussed in \cite{dess1} for the $H\alpha$ line. However, the relative difference
is below 5 \%, thus, these lines are expected to represent $v_{phot}$ with 2 - 4 \% accuracy. 
Motivated by these results, we have selected the He~I $\lambda 5876$ and the Fe~II $\lambda 5169$
features to infer $v_{phot}$ from early-phase (< +20 days) and late-phase spectra of real SNe, 
respectively. 

\begin{figure}
\includegraphics[width=8cm,height=5.5cm]{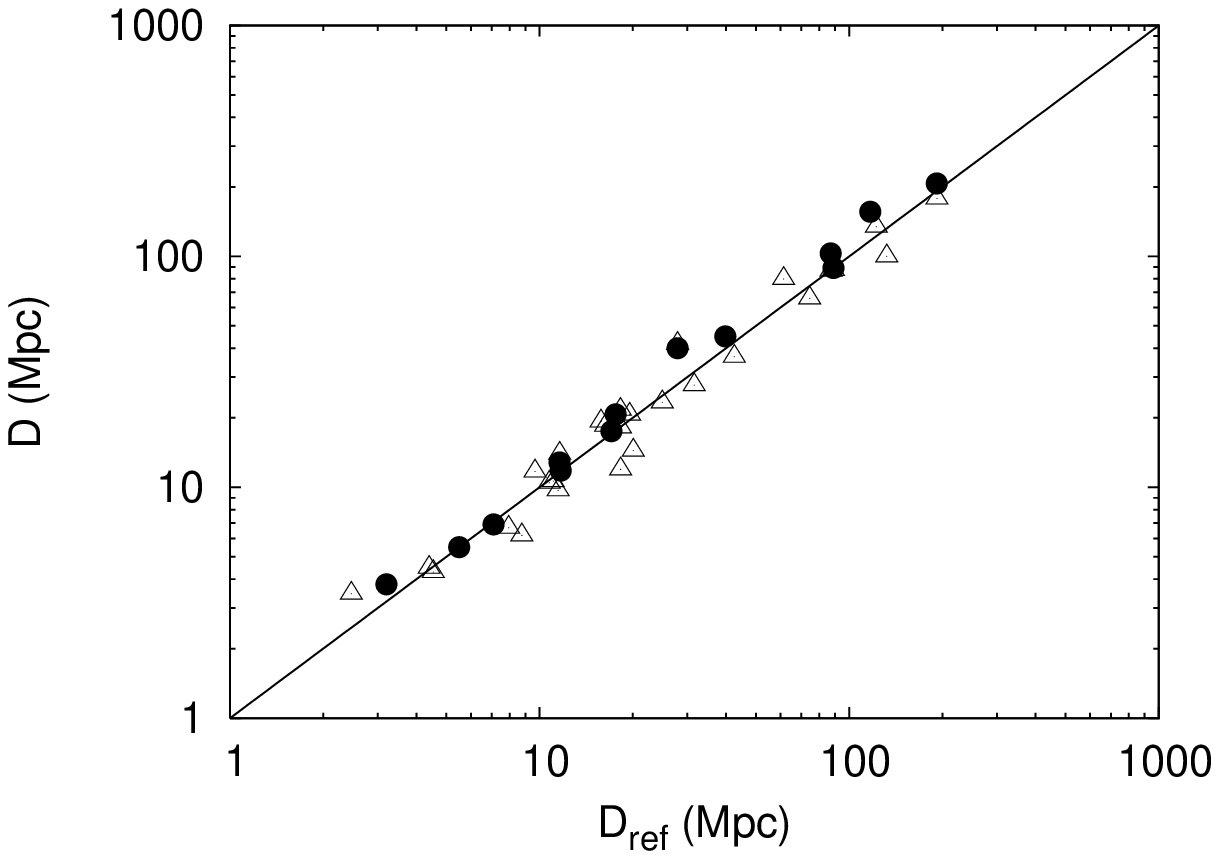}
\includegraphics[width=8cm,height=5.5cm]{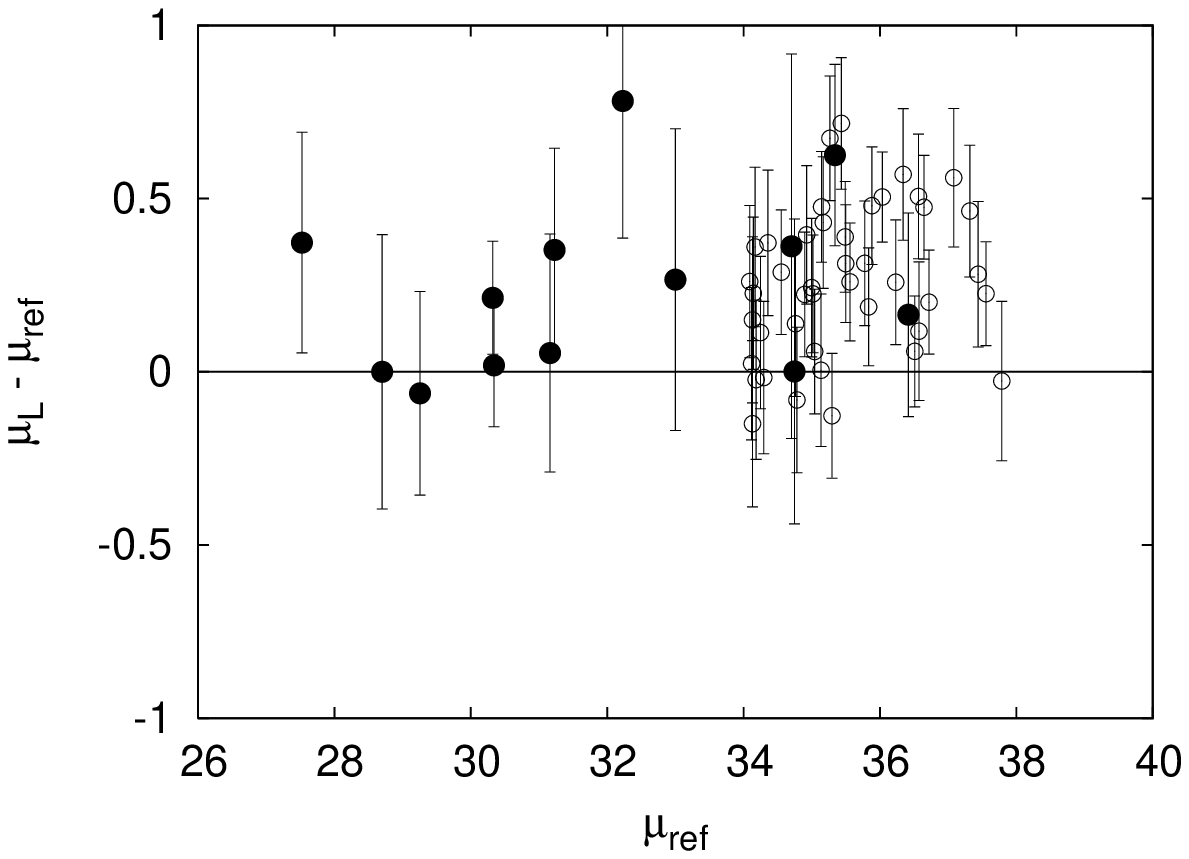}
\caption{Left panel: the comparison of EPM (filled circles) and SCM (open triangles) distances with
the reference distances of the host galaxies. Right panel: residuals of the distance moduli of
Type II-P SNe from EPM (filled symbols) and Type Ia SNe (see text). }
\end{figure}  

In order to apply the method to real SNe, we have collected the available data of Type II-P SNe from
the literature (details and references will be published in a forthcoming paper). 
Eq.1 was fitted to the observed data via least squares using either 
$t$ or $\theta / v_{phot}$ as the independent variable. The two results for $D$ and $t_e$ were averaged to
obtain their final value. Whenever possible, the fit was restricted to data obtained between 
$+5$ -- $+40$ days after explosion, and the angular radii were calculated using the Dessart \& Hillier 
correction factors (see above). In a few cases only late-phase ($t \sim 40 - 60$ days) data were available. 
The Eastman et al. correction factors  were applied for those SNe. 

The EPM distances are plotted against the ``reference'' distances
to their host galaxies (mostly Tully-Fisher or SBF-distances for the nearby ones and Hubble-flow distances
for the more distant ones) in the left panel of Fig.3. As a comparison, the distances coming from the 
,,Standard Candle Method'' (SCM) \cite{hamuy05} for nearly the same observational sample are
also shown. The scattering is very similar for both EPM and SCM.
It is concluded that these two methods provide distances to Type II-P SNe with 
$\sim 15 - 20$ \% accuracy.

The accuracy of the new EPM distances is also similar to that of individual 
SNe Ia distances. This is illustrated in the right panel of Fig.3, where the difference between
the distance moduli of Type II-P SNe (from this paper) and the low-redshift subsample of 
Type Ia SNe (from \cite{riessweb})  are plotted against the reference
distance moduli (adopting $D_{ref} = c z / H_0$ for Type Ia SNe). Again, the scattering
of the data is similar for the two samples. Thus, the concept of EPM combined with the
present knowledge of Type II-P SNe atmospheres may provide consistent and reliable distances,
which may be extended toward higher redshifts in the future. This could be a very important,
independent test of the Type Ia SNe distance scale.
\vskip 0.5cm
This work was supported by Hungarian OTKA Grants No. T 042509 and TS 049872.


\end{document}